\documentstyle[psfig,fleqn]{tp}

\def\etal{{\it et\thinspace al\/}}

\def\be{\begin{equation}}
\def\ee{\end{equation}}
\def\bea{\begin{eqnarray}}
\def\eea{\end{eqnarray}}
\def\Mpc{$h^{-1}$~{\rm Mpc}}
\def\hmpc{$h$~{\rm Mpc$^{-1}$}}

\begin{document}
\twocolumn[

\title{The Power Spectrum of Matter}

\author{J. Einasto\\
{\it Tartu Observatory, EE-2444 T\~oravere, Estonia}}
\vspace*{16pt}   

ABSTRACT.\ We calculate the mean power spectrum of galaxies using
published power spectra of galaxies and clusters of galaxies.  The
mean power spectrum of cluster samples as well as APM 3-D, IRAS QDOT,
and SSRS+CfA2 galaxy samples has a relatively sharp maximum at
wavenumber $k=0.05 \pm 0.01$~\hmpc, followed by an almost exact
power-law spectrum of index $n\approx -1.9$ toward smaller scales.
The power spectrum found from APM 2-D galaxy distribution and from
LCRS and IRAS 1.2 Jy surveys is flatter around the maximum.  We show
that power spectra of galaxies and matter are similar in shape for a
wide class of assumptions, the bias parameter of galaxies relative to
matter is fixed by the fraction of clustered matter associated with
galaxies; and find $b_{gal}=1.3 \pm 0.1$. We compare the empiric power
spectrum of matter with analytical power spectra and show that the
primordial power spectrum has a break in amplitude and a spike if
presently available galaxy and cluster samples represent the true mass
distribution of the Universe.  \endabstract]

\markboth{J. Einasto}{The Power Spectrum of Matter}

\small

\section{The power spectrum of galaxies and clusters}

There exist a large body of observational determinations of the power
spectrum of galaxies and clusters of galaxies. In this talk I shall
review the determination of the mean galaxy power spectrum (Einasto
\etal 1998a, hereafter E98a).  The mean galaxy power spectrum shall be
reduced to that of matter (Einasto \etal 1998b, E98b). This
semi-empirical matter power spectrum shall be used to determine the
primordial power spectrum (Einasto \etal 1998c, E98c).

\begin{figure*} 
\vspace*{6.5cm} 
\caption{Power spectra of galaxies and clusters of galaxies scaled to
match the amplitude of the 2-D APM galaxy power spectrum (E98a).
Spectra are shown as smooth curves and are designated as follows:
ACO-E and ACO-R -- Abell-ACO clusters (Einasto \etal 1997a, Retzlaff
\etal 1998); APM-T -- APM clusters (Tadros \etal 1998); APM-TE -- APM
galaxies (Tadros \& Efstathiou 1996); APM-P-GB -- spectra derived from
2-D distribution of APM galaxies (Peacock 1997, Gazta\~naga \& Baugh
1997); IRAS-P -- IRAS galaxies (Peacock 1997); CfA2 -- SSRS-CfA2~130
Mpc galaxy survey (da Costa \etal 1994); LCRS -- LCRS survey (Landy
\etal 1996); $P(k)_{mean}$ indicates the mean power spectrum $P_{HD}$
for high-density regions; the power spectrum for medium-density
regions, $P_{MD}$, is identified with the spectrum APM-P-GB.  The mean
error of the mean spectrum is 11~\%, for individual samples it varies
between 23 and 48~\%.  } 
\includegraphics{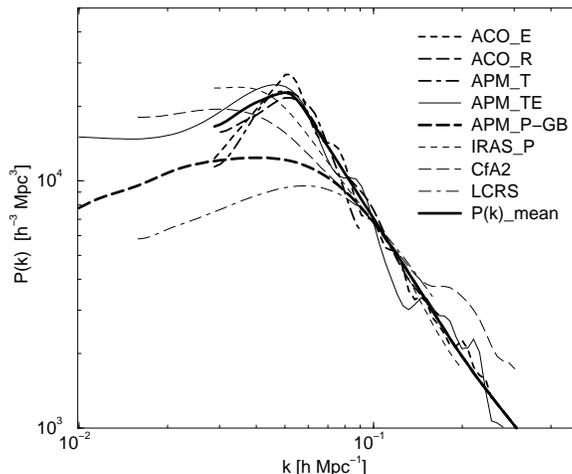}
\label{figure1}
\end{figure*}

Recent determinations of power spectra for large galaxy and cluster
samples are plotted in Figure~1.  The compilation in based on summary
by E98a. Spectra are shifted in amplitude to match the amplitude of
the power spectrum of APM galaxies on wavenumber $k=0.1$~\hmpc.  The
APM galaxy spectrum is a reconstruction of the 3-dimensional spectrum
based on deep 2-dimensional distribution of over 2 millions of
galaxies, thus the cosmic error is smaller here than in available 3-D
surveys. The APM galaxy spectrum is also free of redshift
distortions. We see that after vertical scaling there is little
scatter between individual determinations of power spectra on medium
and small scales.  On large scales around the maximum the scatter is
much larger.

We have formed two mean power spectra.  One spectrum is based on
samples having power spectra with a high amplitude near the maximum.
Such samples are Abell-ACO and APM cluster surveys, the redshift
survey of APM galaxies, and the SSRS-CfA2-130 galaxy survey. These
samples cover large regions in space where both high- and
medium-density regions are well represented, thus we use the notation
$P_{HD}$ for this mean power spectrum (HD for high-density). This mean
power spectrum has a relatively sharp maximum at $k=0.05 \pm
0.01$~\hmpc, followed by an almost exact power-law spectrum of index
$n\approx -1.9$ toward smaller scales.

The other mean spectrum is based on samples which have a power
spectrum with a shallower turnover; such samples are the LCRS survey,
IRAS QDOT galaxies, and the APM 2-D sample of galaxies. In LCRS and
IRAS QDOT surveys medium-density regions are well present but not
regions of highest density (very rich superclusters).  We use the
notation $P_{MD}$ for this mean power spectrum (MD for
medium-density). On medium and small scales it coincides with the
previous spectrum, but it has a maximum of lower amplitude.

Presently it is not clear, whether the difference between these two
spectra is real or partly due to some artifacts of data handling.
Several arguments suggest that there exist real differences between
power spectra of various samples.  All samples which have a high
amplitude of the power spectrum near the maximum are deep fully 3-D
samples. In contrast, samples with a shallower power spectrum are not
so deep, or do not contain very rich superclusters as the LCRS
sample. In IRAS QDOT sample galaxies in rich superclusters were
removed (Tadros \& Efstathiou 1995).

On the other hand, some artifacts of data reduction or influence of
the sample selection and/or geometry are not excluded. For instance,
the Las Campanas survey is not a fully 3-dimensional sample, it is
made in narrow strips which may smooth out sharp features of the power
spectrum near the maximum.  A curious fact is the reconstruction of the
3-D spectrum from APM 2-D galaxy distribution which has near the
maximum a lower amplitude as expected from real 3-D observations of
galaxies of the same sample. This difference is not explained yet, it
may be due to some problems with the reconstruction of the 3-D power
spectrum from 2-D data.

The difference between two mean power spectra can be considered as the
combined result of the cosmic scatter and our ignorance of all details
of the data reduction.

\section{The reduction of galaxy power spectrum to matter}

To compare the observed power spectrum with theoretical spectra the
galaxy spectrum must be reduced to matter. 

Differences between the distribution of galaxies and matter are due to
the gravitational character of the evolution of the Universe. As shown
already by Zeldovich (1970), the evolution of under- and over-dense
regions is different. Matter flows away from low-density regions
toward high-density ones until it collapses. In order to form a galaxy
or a system of galaxies, the mean density of matter in a sphere of
radius $r$ must exceed the mean density by a factor of 1.68 (Press \&
Schechter 1974), the radius $r$ determines the mass of the contracting
object. Thus in low-density regions (voids) galaxies are absent, but
the gravity is unable to evacuate voids completely -- there exists
primordial matter in voids. Visible matter is concentrated together
with dark matter in a web of galaxy filaments and superclusters
(Zeldovich, Einasto \& Shandarin 1982, Bond, Kofman \& Pogosyan 1996).

\begin{figure} 
\vspace{4.5cm} 
\caption{The biasing parameter as a function of the wavenumber $k$
for 2-D simulation, determined for all galaxies (threshold density
$\rho_0=1$ in units of the mean density), galaxies in high-density
regions ($\rho_0= 2$), and galaxies in clusters ($\rho_0=5$).  }
\includegraphics{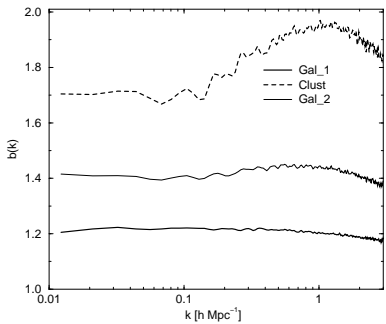}
\label{figure2}
\end{figure}

These considerations show that model particles can be divided into two
populations, the unclustered population in voids, and the clustered
population in high-density regions. The last population is associated
with galaxies including DM halos of galaxies and clusters of
galaxies. To get the clustered population one has to exclude the
population of low-density particles using a certain threshold density,
$\rho_0$, which divides the matter into the matter in voids and the
clustered matter.  Hydrodynamical simulations by Cen \& Ostriker
(1992, 1998) show that the overall mean density is a good
approximation to the threshold density. In determining the density
field we use smoothing on scales comparable to the size of actual
systems of galaxies ($\sim 1$~\Mpc).

Analytical calculations and numerical simulations show that the
exclusion of matter from low-density regions rises the amplitude of
the power spectrum but not its shape (E98b). Power spectra of galaxies
and matter are related as follows:
$$
P_{gal}(k) = b^2 P_{m}(k),
\eqno(1)
$$
where the bias factor $b$ is expressed by the fraction of matter in
the clustered population associated with galaxies, 
$$
b=1/F_{gal}.
\eqno(2)
$$
Actually the biasing parameter $b$ is a slow function of the
wavenumber $k$, but in the range of scales of interest for the present
study it is practically constant.  The biasing parameter $b(k)$ is
shown in Figure~2 for a 2-D model, it is found by comparing power
spectra of all particles (matter) and clustered particles associated
with galaxies using different threshold density $\rho_0$ (E98b).

\begin{figure} 
\vspace{4.5cm} 
\caption{The fraction of matter associated with galaxies, $F_{gal}$,
as a function of time, measured through the $\sigma_8$ parameter
(curved lines). Thick straight line shows the relation equation~(3); open
circle notes the observed value of $\sigma_8$.  }
\includegraphics{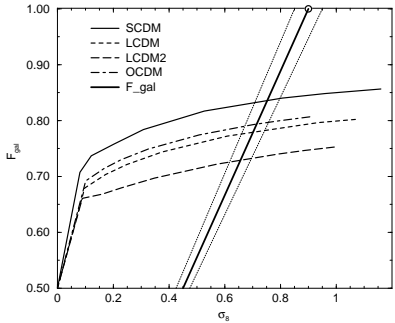}
\label{figure3}
\end{figure}

\begin{figure*}
\vspace*{4.5cm}  
\caption{The semi-empirical matter power spectra compared with
theoretical and primordial power spectra for mixed DM models. Left:
present power spectra; right: primordial spectra. Solid bold line
shows the matter power spectrum found for regions including rich
superclusters, $P_{HD}(k)$; dashed bold line shows the power spectrum
of matter $P_{MD}(k)$ for medium dense regions in the Universe. On
small scales observed power spectra are corrected for non-linear
effects.  Model spectra with $\Omega_0=0.9, \dots ~0.3$ are plotted
with solid lines; for clarity models with $\Omega_{0} = 1.0$ and $0.5$
are drawn with dashed lines.  Primordial power spectra are shown for
the power spectrum $P_{HD}(k)$; they are reduced to scale-free
spectrum, $P(k) \sim k$.  }  
\includegraphics{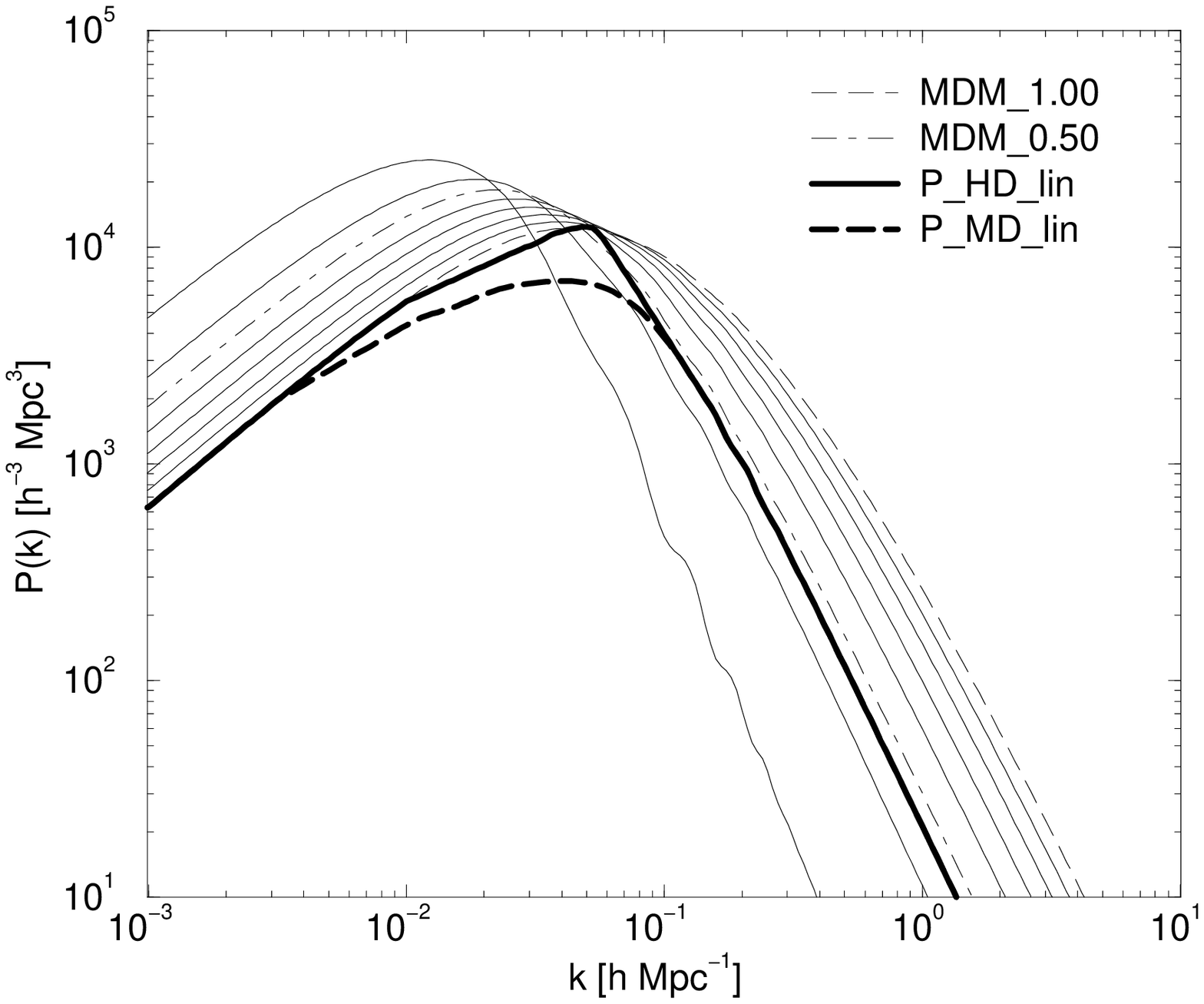}
\includegraphics{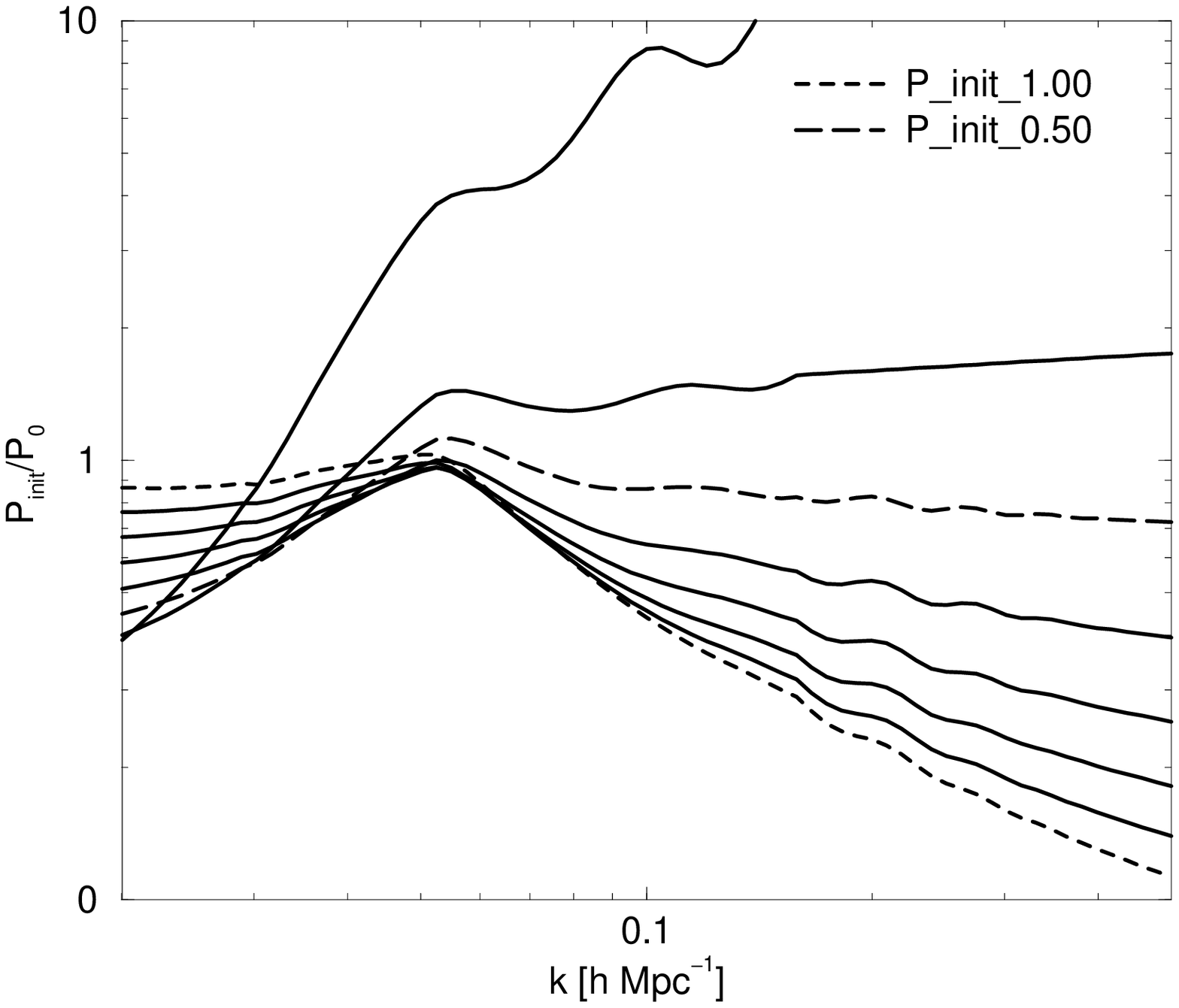}
\label{figure4}
\end{figure*}

The fraction of matter in the clustered population can be determined
from numerical simulations of the void evacuation. This has been done
for several models: the standard CDM model (SCDM, Hubble constant
$h=0.5$), model with cosmological constant (LCDM, $h=0.7$,
$\Omega_0=0.3$), and open model (OCDM, $h=0.7$, $\Omega_0=0.5$).  In
Figure~3 we show the increase of the fraction of matter associated
with galaxies for these models (for LCDM 2 models with different
realization are given).  The epoch of simulation is measured in terms
of $\sigma_8$, rms density fluctuations in a sphere of radius
$r=8$~\Mpc.  From model data alone it is impossible to determine
$\sigma_8$ which corresponds to the present epoch. The value of this
parameter is known from observations for galaxies. It is found by the
integration of the observed mean power spectrum of galaxies; we get
$(\sigma_8)_{gal}=0.89 \pm 0.05$.  This value is equal to $\sigma_8$
for matter, if all matter were associated with galaxies
($F_{gal}=1$). In reality they are different and related via an
equation similar to (1) since $\sigma_8^2$ is proportional to the
amplitude of the power spectrum:
$$
(\sigma_8)_m = F_{gal}(\sigma_8)_{gal}.
\eqno(3)
$$  
This relationship is plotted in Figure~3 by a straight line.  The
intersection of this line with curves $F_{gal}$ vs. $\sigma_8$ for
models yields values of both parameters which correspond to the
present epoch.  We get $\sigma_8=0.68 \pm 0.06$ for matter;
$F_{gal}=0.75 \pm 0.05$, and the biasing parameter of galaxies in
respect to matter $b_{gal}=1.3 \pm 0.1$ (E98b).

\section{The primordial power spectrum}

The final step in our study is the comparison of the power spectrum of
mass with theoretical power spectra for models of structure formation.
We use models with cold dark matter (CDM) and a mixture of cold and
hot dark matter (MDM) in spatially flat models. We derive the transfer
functions for a set of cosmological parameters. 

Models are normalized on large scale by four-year COBE normalization
(Bunn \& White 1997); the density of matter in baryons is taken
$\Omega_{b} = 0.04$ (in units of the critical density); and the Hubble
parameter $h = 0.6$.  The cosmological constant was varied between
$\Omega_{\Lambda} = 0$ and $\Omega_{\Lambda} = 0.8$. In mixed dark
matter models the density of hot DM was fixed, $\Omega_{\nu}=0.1$, cold
DM density was chosen to get a spatially flat model.

Analytical power spectra for MDM models are plotted in the left panel
of Figure~4 together with the semi-empirical matter power spectra,
$P_{HD}(k)$ and $P_{MD}(k)$. MDM models fit the semi-empirical matter
power spectrum better than other models.  In the right panel we show
the initial power spectrum,
$$
P_{init}(k) = P(k)/T^{2}(k),
\eqno(4)
$$
compared with the scale-free primordial power spectrum, $P_0(k)  \sim k$;
here $T(k)$ is the transfer function. The initial power spectrum is
plotted only for spectrum $P_{HD}(k)$ which corresponds to high-density
regions.

The main feature of primordial power spectra is the presence of a
spike at the same wavenumber as that of the maximum of the observed
power spectrum. Primordial spectra also have a break, i.e.  their
amplitudes on small scales are different from the amplitudes on large
scales.  The shape of the primordial spectrum varies with the
cosmological constant. The primordial power spectrum which is found from
the shallower spectrum $P_{MD}(k)$ has no sharp peak but the break is
similar to that for the spectrum $P_{HD}$. 

The main conclusion from the present analysis is that it is impossible
to avoid a break and/or spike in the primordial power spectrum, if
presently available cluster and galaxy data represent a fair sample of
the Universe. Clusters of galaxies cover a much larger volume in space
than galaxies, thus cluster samples are presently best candidates for
the fair sample. However, this conclusion is tentative. New very deep
surveys of galaxies now in progress will specify the power spectrum on
large scales more exactly and yield a better estimate of the
primordial power spectrum.  Presently we can say that the possibility
of a broken scale and peaked initial power spectrum has to be taken
seriously.

\section*{Acknowledgments}

I thank H. Andernach, F. Atrio-Barandela, R. Cen, M. Einasto,
M. Gramann, A. Knebe, V. M\"uller, I. Suhhonenko, A. Starobinsky,
E. Tago and D. Tucker for fruitful collaboration and permission to use
our joint results in this talk.  This study was supported by the
Estonian Science Foundation.

\end{document}